\documentclass[aps,prb,floatfix,twocolumn]{revtex4}
\usepackage{graphicx,amsmath,amssymb,subfigure,epsfig,psfrag,color,verbatim}
\providecommand{\mean}[1]{\left\langle #1 \right\rangle}

\newcommand{\myhrulefill}{}
\begin{document}
\title{Bond-Propagation Algorithm for Thermodynamic Functions in General 2D Ising Models}
\author{Y.~L.~Loh}
\author{E.~W.~Carlson}
\affiliation{Department of Physics, Purdue University, West Lafayette, IN  47907 }
\author{M.~Y.~J.~Tan}
\affiliation{Department of Physics, University of Illinois at Urbana-Champaign, Urbana, IL  61801}
\date{\today}
\begin{abstract}
Recently, we developed and implemented the bond propagation algorithm for calculating the partition function and correlation functions of random bond Ising models in two dimensions.\cite{loh2006} The algorithm is the fastest available for calculating these quantities near the percolation threshold.  In this paper, we show how to extend the bond propagation algorithm to directly calculate thermodynamic functions by applying the algorithm to derivatives of the partition function, and we derive explicit expressions for this transformation.  We also discuss variations of the original bond propagation procedure within
the larger context of $Y\nabla Y$-reducibility and discuss the relation of this
class of algorithm to other algorithms developed for Ising systems.  
We conclude with a discussion on the outlook for applying similar algorithms
to other models.  
\end{abstract}
\maketitle
	
\myhrulefill\section{Introduction}
For nearly 80 years 
Ising models
 have given valuable insight into phase transitions and critical phenomena in magnets, alloys, and many other systems.  Random-bond Ising models (RBIMs) in particular are often used to study frustration and spin-glass behavior, and they are closely related to neural networks and information theory.\cite{nishimoribook}
In Ref.~\onlinecite{loh2006}, we presented a numerically exact
\footnote{{\em I.e}., its accuracy is limited by the precision of the floating-point numbers used in the computation; it suffers from roundoff error but not statistical error like Monte Carlo methods}
algorithm for computing the partition function and correlation functions in a class of 2D Ising models, which works for any planar network of Ising spins with arbitrary bond strengths but without applied fields in the bulk.  Applications include random-bond Ising models (RBIM) (including $\pm K$ disorder, Gaussian disorder, site dilution, and bond dilution) and geometric frustration as in the case of triangular Ising antiferromagnets.
Here, we show how to extend the algorithm to directly calculate thermodynamic quantities such 
as the internal energy.

Since its introduction in 1925, many exact results have been found for regular 2D Ising models
({\em i.e.}, those with translational symmetry in two directions).  
There has been continued interest in random-bond Ising models, which have more complexity because of the need to average not only over all spin configurations, but also over all configurations of bond strengths.  Approximate analytic methods can give misleading or conflicting results, so numerical calculations play an important role in the field.  

Numerical methods for 2D RBIMs have developed along several lines.  
Onsager's original solution using operator techniques generalizes to the fermion-network method of Merz and Chalker\cite{merz}.
The Ising partition function is related to the number of dimer coverings (perfect matchings) of a modified graph, and hence to the problem of computing a determinant or Pfaffian\cite{kasteleyn1963,fisher1966}; this approach was applied to RBIMs by Saul and Kardar\cite{saul1993,saul1994} and by Galluccio
{\em et al}\cite{chineseremainder}.
The fastest of these algorithms\cite{chineseremainder} takes of 
order $O(N^{3/2})$ time for a network of $N$ spins.  
Algorithms in this class have the disadvantage that one must first 
map to fermions or dimers before the model can be solved.  

A different line of development involves the $Y$-$\nabla$ and $\nabla$-$Y$
transformations for the partition functions of Ising  models,\cite{houtappel,syozi}
operating directly in the spin basis, without the need to map to fermion or dimer models.
In a $Y$-$\nabla$ (or star-triangle) transformation, three Ising spins connected by bonds to a
central spin can be converted into the same three Ising spins but without the
central spin, now connected
by mutual bonds, in such a way as to preserve the total partition function of the system.
The reverse $\nabla$-$Y$ transformation can also be done so as to preserve the 
partition function.  
For networks which are $Y\nabla Y$-reducible, 
one can then compute the partition function $Z$ at a given temperature using
a suitable sequence of such transformations. 
Colbourn \emph{et al.} \cite{colbourn1995} suggested using the Feo-Provan\cite{feo1993} $Y\nabla Y $ reduction method for general networks, which takes $O(N^2)$ time where $N$ is the number of nodes in the network (i.e., spins). 
Frank and Lobb invented the bond propagation algorithm\cite{lobb} (a form of $Y\nabla Y $ reduction) for 2D resistor networks which takes $O(N^{3/2})$ time for square lattices 
(grid graphs).\footnote{The algorithms in Refs.~\onlinecite{truemper1989,gitlerthesis,gitlerpreprint} also take $O(N^{3/2})$ time.}
They also suggested an extension to Ising models.    
Recently, we developed and implemented the Ising bond propagation algorithm.\cite{loh2006}  
The algorithm executes in $O(N^{3/2})$ time for most planar networks of interest, 
and in $N \ln N$ time for dilute models near percolation,\cite{lobb}
making it the fastest method for computing the Ising partition function and correlation
functions near the percolation threshold.
In this paper, we derive the transformations necessary to implement the algorithm
on derivatives of the partition function, allowing for
fast, direct, and exact calculation
of thermodynamic quantities.  We also discuss the relation of this class of algorithm to
graph theory as well as to other methods for Ising systems.

The outline of this paper is as follows.  
In Sec.~\ref{s:models} we introduce the models to which the
algorithm is applicable.
In Sec.~\ref{s:transformations}
we derive the network transformations, most notably the $Y$-$\nabla$ and $\nabla$-$Y$ transformations, that are the building blocks of the bond-propagation algorithm (BPA)
for both resistor networks ({\em i.e.}, Gaussian models) and the partition function of Ising models,
including new forms of the transformations which have improved numerical stability
in the latter case. 
In Sec.~\ref{s:bondprop} we describe some variations of the algorithm.
In Sec.~\ref{s:jeremy} we show that the BPA can be used to compute derivatives of the Ising partition function directly, and we derive these transformations for the first derivative with respect to temperature. 
In Sec.~\ref{s:discussion} we discuss practical issues such as the strengths and limitations
of the BPA, as well as the outlook for application of similar algorithms to other models.

\myhrulefill\section{Models}\label{s:models}
$Y\nabla Y$ reduction is applicable to both Gaussian models (such as resistor networks) and Ising models.  
We define a `Gaussian model' in statistical mechanics as one whose action $S(\{\phi\})$ is bilinear in the fields $\phi$, which may be real, complex or Grassmannian,
	\begin{align}
	e^{ S(\{\phi\}) } &= \exp \sum_{ij} K_{ij} \phi_i^* \phi_j . \label{e:linearmodelaction}
	\end{align}
This action also applies to dynamical models that are diagonal in frequency space, such as electrical $LCR$ networks involving inductors, capacitors and/or resistors where the $K_{ij}$ represent admittances (complex conductances), and non-interacting bosons, fermions, phonons or magnons on lattices with hopping amplitudes $K_{ij}$.

An Ising model action is a quadratic form in spin variables $\sigma_i=\pm 1$ with a symmetric kernel $K_{ij}$, 
	\begin{align}
	e^{ S(\{\sigma\}) }
	= e^{ -\beta \mathcal{H} (\{\sigma\}) }
	&= \exp \sum_{\mean{ij}} K_{ij} \sigma_i \sigma_j
	\label{e:isingmodelaction}
	\end{align}
where $\mean{ij}$ indicates a sum over pairs of spins, $K=\beta\widetilde{K}$ are dimensionless Ising couplings, $\widetilde{K}$ are the physical couplings with dimensions of energy, and $\beta=1/T$ is the inverse temperature.  
This includes various types of random-bond Ising models (RBIMs) such as those with `binary' couplings ($K_{ij}=\pm K_0$), Gaussian couplings, and zero couplings (dilute Ising models).  Such models are often used to study spin glasses and dilute magnets.

In the Ising system, our goal is to calculate the partition function,
	\begin{align}
	Z &= \sum_{\{\sigma\}} e^{ S(\{\sigma\}) }
	, 
	\end{align}
correlation functions such as
	\begin{align}
	\mean{\sigma_i \sigma_j} &= Z^{-1} \sum_{\{\sigma\}} \sigma_i \sigma_j e^{ S(\{\sigma\}) }
	,
	\end{align}
and/or thermodynamic functions such as
	\begin{align}
	U &= -Z^{-1} \tfrac{dZ}{d\beta}
	= Z^{-1} \sum_{\{\sigma\}} \mathcal{H} (\{\sigma\}) e^{ S(\{\sigma\}) }
	.
	\end{align}
For Gaussian models such as resistor networks,
we aim instead to calculate the effective resistance $R_{ij}$ between two points in the network.  

Computing the partition function of a many-body system is usually a very difficult problem that involves summing over a number of configurations that is exponential in system size.  Even if this is re-cast into the form of a high-temperature or low-temperature series, it is still usually the case that one needs to sum over a large number of graphs, or count a large number of graph embeddings, using some amount of brute force.
Treatments based on successive elimination of spins are 
often limited because integrating out spin degrees of freedom tends to generate 
longer range couplings or 
multispin interactions 
such as $3$-spin couplings.
Another approach is real-space decimation, which is approximate because one has to discard most interactions beyond a certain level of complexity.

There are some nice exceptions: for Gaussian models and Ising models on planar networks,
\footnote{
For Gaussian models such as Eq.~\eqref{e:linearmodelaction}, the partition function $Z$ is related to the determinant $\det K$, and for planar Ising models $Z$ is related to the Pfaffian of the ``Kasteleyn terminal lattice'', both of which may be computed in polynomial time.  It is interesting that these same two models are amenable to the bond-propagation algorithm which we describe in this paper.
}
it is possible to reduce an entire network by successive 
$Y$-$\nabla$ and $\nabla$-$Y$ transformations,
local moves which never induce multispin interactions.  Thus throughout the whole procedure the system can be described just by pairwise interactions; therefore, it is useful to adopt the language of graph theory and represent it by a network, with nodes (vertices), edges (bonds), and edge weights (couplings).
The next section describes transformations that can be used in the reduction of such a `statistical mechanics network'.
  
\myhrulefill\section{Star-Mesh and Mesh-Star Transformations}\label{s:transformations}
\subsection{Gaussian models}
\begin{figure}
\psfrag{J01}{$K_{01}$} \psfrag{J23}{$K_{23}$} \psfrag{J23'}{$K_{23}'$} \psfrag{s1}{$\phi_1$}
\psfrag{J02}{$K_{02}$} \psfrag{J31}{$K_{31}$} \psfrag{J31'}{$K_{31}'$} \psfrag{s2}{$\phi_2$}
\psfrag{J03}{$K_{03}$} \psfrag{J12}{$K_{12}$} \psfrag{J12'}{$K_{12}'$} \psfrag{s3}{$\phi_3$}
\psfrag{s0}{$\phi_0$}
{\centering
{\resizebox*{0.95\columnwidth}{!}{\includegraphics{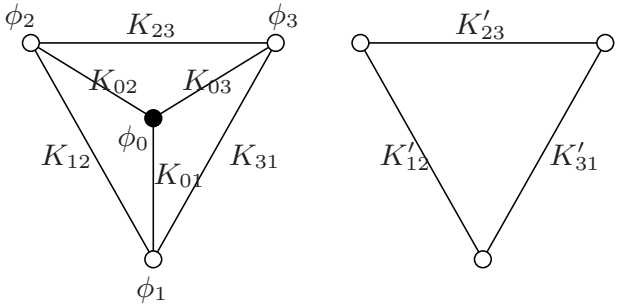}}}
\par}
\caption{\label{f:starmesh2}
Elimination of an $N$-degree node, illustrated for $N=3$.  Diagonal couplings $K_{11},K_{22},K_{33}$ correspond to loops (edges connecting a node to itself) and have been omitted.
}
\end{figure}

For pedagogical reasons, we first review Gaussian models (which include resistor networks as a special case).
We show how to eliminate any node of a Gaussian model via a star-mesh transformation regardless of its coordination number.  Consider the network in Fig.~\ref{f:starmesh2} described by an action $S(\phi_0,\phi_1,\dotsc,\phi_N)$ containing bilinear couplings $K_{ij} \phi^*_i \phi_j$ as well as a constant `free energy' term $F$.  The variable $\phi_0$ can be eliminated by Gaussian integration, leading to a new effective action $S'(\phi_1,\dotsc,\phi_N)$ with parameters $F'$ and $K'_{ij}$:
	\begin{align}
	& \int (d\phi_0)~ 	\exp {S(\phi_0,\phi_1,\dotsc,\phi_N)} \nonumber\\
	&=\int (d\phi_0)~ 	\exp \left( F + \sum_{i=0}^N \sum_{j=0}^N K_{ij} \phi_i^* \phi_j \right) \nonumber\\
	&=\exp \left( F + \zeta \ln \frac{-1}{K_{00}} 
			+ \sum_{i=1}^N \sum_{j=1}^N \left( K_{ij} - \frac{K_{i0}K_{0j}}{K_{00}} \right) \phi_i^* \phi_j \right) \nonumber\\
	&=\exp \left( F' 
			+ \sum_{i=1}^N \sum_{j=1}^N K'_{ij}  \phi_i^* \phi_j \right)  \nonumber\\
	&= \exp {S'(\phi_1,\dotsc,\phi_N)}  
	\label{e:gaussianstarmeshderivation}
	\end{align}
where $\int (d\phi) \equiv \int \frac{d(\Re \phi) d(\Im \phi)}{\pi}$ and $\zeta=1$ in the case where $\phi$ are complex variables; similar expressions exist for real $\phi$ and Grassmannian $\phi$.  The integration generates additive changes to the free energy, 
 $\delta F=F'-F$, and to the couplings, $\delta K_{ij}=K'_{ij}-K_{ij}$:
	\begin{align}
	\delta F &= \zeta \ln \frac{-1}{K_{00}} ,\label{e:gaussianstarmesh1} \\
	\delta K_{ij} &= - \frac{K_{i0}K_{0j}}{K_{00}}, \quad i,j=1,2,3,\dotsc,N. \label{e:gaussianstarmesh0}
	\end{align}
Eq.~\eqref{e:gaussianstarmesh0} contains only algebraic operations---addition, multiplication and division---and it is homogeneous, i.e., invariant under multiplication of all $K$'s by the same constant.  In fact, Eq.~\eqref{e:gaussianstarmesh0} corresponds to row and column subtractions on the the matrix $K_{ij}$;
the elimination of $\phi_0$ by Gaussian integration is closely related to the procedure of Gauss elimination in linear algebra.

The above derivation is valid even if the variables $\phi_1 \dotsc \phi_N$ are coupled to external fields or additional variables; these extra terms simply cancel out on both sides of 
Eq.~\eqref{e:gaussianstarmeshderivation}.

\begin{figure}
\psfrag{J01}{$K_{01}$} \psfrag{J23}{$K_{23}$} \psfrag{J23}{$K_{23}'$} \psfrag{s1}{$\phi_1$}
\psfrag{J02}{$K_{02}$} \psfrag{J31}{$K_{31}$} \psfrag{J31}{$K_{31}'$} \psfrag{s2}{$\phi_2$}
\psfrag{J03}{$K_{03}$} \psfrag{J12}{$K_{12}$} \psfrag{J12}{$K_{12}'$} \psfrag{s3}{$\phi_3$}
{\centering
{\resizebox*{0.95\columnwidth}{!}{\includegraphics{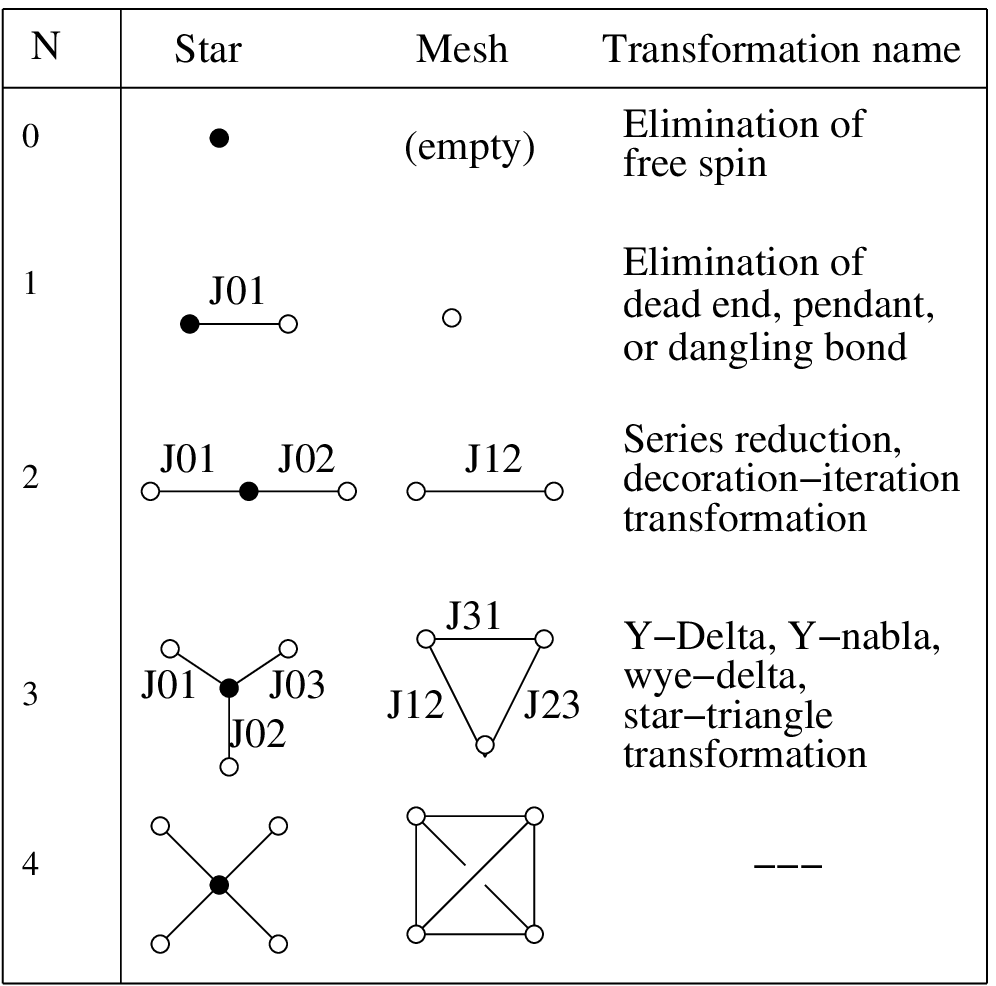}}}
\par}
\caption{\label{f:starmesh}
Star-mesh transformations for various $N$.
Circles represent nodes and lines represent edges; the filled node is the one being eliminated.
}
\end{figure}

Node elimination corresponds to the ``star-mesh transformations'' illustrated in Fig.~\ref{f:starmesh} and defined as follows.
An \emph{$N$-degree star-mesh transformation} eliminates a node of coordination number $N$, introducing $N(N-1)/2$ edges between its neighbors.
A mesh-star transformation is the inverse, which does not always exist.

For Gaussian models the star-mesh formula is 
	\begin{align}
	K'_{ij} = - K_{i0}K_{0j}/K_{00}, \quad i,j=1,2,3,\dotsc,N.
	\label{e:gaussianstarmesh}
	\end{align}
In the special case of a resistor network, the off-diagonal elements of the kernel are equal to the conductances, $K_{ij}=G_{ij}$, and the diagonal elements are given by a sum rule $K_{ii}=-\sum_{j=1}^N G_{ij}$, so the star-mesh formula becomes $G'_{ij}=G_{i0}G_{0j} / \sum_{j=1}^N G_{0j}$.  
For $N=2$, this corresponds to the familiar 
series
reduction of conductances.
For $N=3$, it reduces to the $Y$-$\nabla$ transformation,
	\begin{align}
	G_{12}' &= G_{01} G_{02} / (G_{01} + G_{02} + G_{03})  \qquad\text{(cycl.)},
	\label{e:conductanceynabla}
	\end{align}
where `cycl.' indicates additional equations in which the indices $1,2,3$ are cyclically permuted.  Eqs.~\eqref{e:conductanceynabla} can be inverted to give the $\nabla$-$Y$ transformation, which is most easily written in terms of the resistances $R=1/G$:
	\begin{align}
	R_{01} &= R'_{31} R'_{12} / (R'_{31} + R'_{12} + R'_{23})  \qquad\text{(cycl.)}.
	\label{e:resistancenablay}
	\end{align}
The equations for node-voltage analysis of a planar electrical network are the same as those for loop-current analysis of the dual network, which is obtained by interchanging node voltages $V_i$ with loop currents $I_\mu$, and conductances $G_{ij}$ joining adjacent nodes with resistances $R_{\mu\nu}$ separating adjacent loops.  Thus the similarity between Eqs.\eqref{e:conductanceynabla} and \eqref{e:resistancenablay} is no coincidence.

For computer implementation it may be preferable or necessary to supplement the above formulas by formulas for special cases\cite{lobb}, involving zero couplings (opens) or infinite couplings (shorts).
Assuming that floating-point overflows never occur, the following rules are sufficient:
	\begin{align}
	\mathtt{YDelta} (\infty, G_1, G_2) &= (0, G_2, G_1) \\
	\mathtt{DeltaY} (0, G_1, G_2) &= (\infty, G_2, G_1) \\
	\mathtt{DeltaY} (\infty, G_1, G_2) &= (G_1+G_2, \infty, \infty) .
	\label{e:conductancespecialcases}
	\end{align}
The $G_{ij}$ can be complex numbers representing admittances at a particular frequency, in which case the transformations involve complex arithmetic.

\subsection{Ising models}
For Ising networks, one may attempt to derive a star-mesh transformation in a similar way,
integrating out a spin $\sigma_0$ to generate an effective action $S'$ with  parameters $F'$ and $K'$ to be determined
(see Fig.~\ref{f:starmesh}): 
	\begin{align}
	\sum_{\sigma_0=\pm 1} 	e^ {S(\sigma_0,\sigma_1,\dotsc,\sigma_N)} 
	&= e^ {S'(\sigma_1,\dotsc,\sigma_N)}
	\quad\forall \sigma_1 \dotsc \sigma_N
	\\ 
	\therefore
	\sum_{\sigma_0=\pm 1} 	e^{  F + \sum_{i=1}^{N} K_{0i} \sigma_0 \sigma_i  }
	&=e^{  F' + \sum_{i=1}^{N} \sum_{j=i+1}^{N} K'_{ij}  \sigma_i \sigma_j  } 
	.
	\end{align}
Because $\sigma_1,\dotsc,\sigma_N$ can each take the values $\pm 1$, there are $2^N$ equations that need to be satisfied.  Due to spin-flip symmetry, only $2^{N-1}$ of these are independent.  
For $N\geq 4$ the system of equations is overdetermined, with no solution.  For example, for $N=4$, there are 8 independent equations in 7 unknowns $F'$, $K'_{12}$, $K'_{13}$, $K'_{14}$, $K'_{23}$, $K'_{34}$, $K'_{34}$; the true effective action $S'$  contains a 4-spin interaction of the form $K'_{1234} \sigma_1\sigma_2\sigma_3\sigma_4$ that does not fit within the network formalism.  
It is futile to keep track of such multispin terms because their number can grow to $O(2^N)$.

For $N=3$, however, there are 4 independent equations in 4 unknowns $F'$, $K'_{12}$, $K'_{13}$, $K'_{23}$,
so a $Y$-$\nabla$ transformation exists\cite{houtappel} such that
	\begin{align}
	& \sum_{\sigma_0}~ 	e^{F + K_{01} \sigma_0\sigma_1 + K_{02} \sigma_0\sigma_2 + K_{03} \sigma_0\sigma_3} \nonumber\\
	&=2e^F  \cosh (K_{01} \sigma_1 + K_{02} \sigma_2 + K_{03} \sigma_3) \nonumber\\
	&=e^{F' + K'_{23} \sigma_2\sigma_3 + K'_{31} \sigma_3\sigma_1 + K'_{12} \sigma_1 \sigma_2 }.
	\end{align}
Writing  $\delta F=F'-F$ and the new couplings $K'_{ij}$ in terms of the old couplings $K_{i0}$,
	\begin{align}
	&
		2\cosh \big(
			  K_{01} \sigma_1
			+ K_{02} \sigma_2
			+ K_{03} \sigma_3
		\big)
		\nonumber\\
	&=
		e^{
				\delta F
			+	K'_{23} \sigma_2 \sigma_3
			+	K'_{31} \sigma_3 \sigma_1
			+	K'_{12} \sigma_1 \sigma_2
		}
	\qquad\forall \sigma_1,\sigma_2,\sigma_3
	\label{e:equatingfirstterms}
		\nonumber\\
	\therefore&
	\begin{cases}
	a_0=2\cosh (+ K_{01} + K_{02} + K_{03}) = e^{ \delta F + K'_{23} + K'_{31} + K'_{12} } \\
	a_1=2\cosh (- K_{01} + K_{02} + K_{03}) = e^{ \delta F + K'_{23} - K'_{31} - K'_{12} } \\
	a_2=2\cosh (+ K_{01} - K_{02} + K_{03}) = e^{ \delta F - K'_{23} + K'_{31} - K'_{12} } \\
	a_3=2\cosh (+ K_{01} + K_{02} - K_{03}) = e^{ \delta F - K'_{23} - K'_{31} + K'_{12} } \\
	\end{cases}
		\nonumber\\
	\therefore&
	\begin{cases}
	\delta F = \ln a_0 {}^{1/4} a_1 {}^{1/4} a_2 {}^{1/4} a_3 {}^{1/4} , \\
	K'_{23} =  \ln a_0 {}^{1/4} a_1 {}^{1/4} a_2 {}^{-1/4} a_3 {}^{-1/4} \quad\text{(cycl.)}\\
	\end{cases}
	\end{align}
where $a_i$ are functions of $K_{0i}$ as above.

For computer implementation it is preferable to write the Ising action as
	\begin{align}
	e^ { S(\{\sigma\}) } 
	&= \text{const} \cdot z \prod_{\mean{ij}} k_{ij} {}^{(1-\sigma_i\sigma_j)/2}
	\label{e:isingactionalternative}
	\end{align}
where $z=e^F$ and $k_{ij}=e^{-2 K_{ij}}$.  
In this representation the $Y$-$\nabla$ transformation is 
	\begin{align}
	\mathtt{YDelta}(k_1, k_2, k_3) &= (k_{23}, k_{31}, k_{12}; \delta z) \quad\text{where} \nonumber\\
	\delta z &= 1 + k_1 k_2 k_3 \nonumber\\
	z_1 &= k_1 + k_2 k_3 \qquad\text{(cycl.)} \nonumber\\
	b &= \sqrt{z_1 z_2 z_3/\delta z} \nonumber\\
	k_{23} &= b/z_1 \qquad\text{(cycl.)} 
	\label{e:isingynablaalgebraic}
	\end{align}
where we have written $k_{0i}\equiv k_i$ and $\delta z=e^{\delta F}$ (and omitted primes on $k_{23}$ etc.). 
The transcendental functions (exp, log, cosh) have been replaced by 
algebraic functions ($+$, $\times$, $\div$, $\sqrt{~}$), which are quicker to evaluate and have fewer complications arising from multivaluedness.  Furthermore, infinite ferromagnetic couplings $K=\infty$ (`shorts' in the language of Ref.~\onlinecite{lobb}) are readily represented by $k=0$.

The $\nabla$-$Y$ transformation is equivalent to a $Y$-$\nabla$ transformation in the dual representation using Syozi's `cyclic change of lattices' \cite{syozi}:
	\begin{align}
	\mathtt{DeltaY}(k_{23}, k_{31}, k_{12}) &= (k_1, k_2, k_3) \quad\text{where} \nonumber\\
	p_1 &= \mathtt{dual} (k_{23}) \quad\text{(cycl.)}, \nonumber\\
	(p_{23}, p_{31}, p_{12}) &= \mathtt{YDelta} (p_1, p_2, p_3), \nonumber\\
	k_1 &= \mathtt{dual} (p_{23}) \quad\text{(cycl.)}. 
	\label{e:syozi}
	\end{align}
where the duality relation is the M\"obius transformation $\mathtt{dual}(k)=\frac{1-k}{1+k}$.
However, this approach does not give an expression for the free energy change 
$\delta F$,\cite{houtappel} 
and we have found that a direct implementation of Eq.~\eqref{e:syozi} is susceptible to roundoff error.  The scheme below has better numerical performance:
	\begin{align}
	x_1 &= 2 k_{31} k_{12} (1 - k_{23} {}^2) \quad\text{(cycl.)} \nonumber\\
	y_1 &= 1 + k_{31} {}^2 k_{12} {}^2 - k_{12} {}^2 k_{23} {}^2 - k_{23} {}^2 k_{31} {}^2 \quad\text{(cycl.)} \nonumber\\
	v &= \sqrt{y_1 {}^2 - x_1 {}^2} \nonumber\\
	k_1 &= x_1 / (y_1 + v) 	\text{~~~or~~~} (y_1 - v) / x_1
	\quad\text{(cycl.)} \nonumber\\
	\delta z &= 1 + k_1 k_2 k_3
	\label{e:isingnablayalgebraic}
	\end{align}
The two expressions for $k_1$ are formally equivalent; the choice depends on which one is numerically more stable (i.e., avoids subtraction of similar quantities).

Eqs.~\eqref{e:isingynablaalgebraic} and \eqref{e:isingnablayalgebraic}  are preferable to those in Refs.~\onlinecite{colbourn1995} and \onlinecite{loh2006}, because they each contain only one square root, and have been optimized to reduce roundoff error.  Both transformations are 1-to-2 mappings by virtue of the square roots,
 whose signs can be chosen arbitrarily.\footnote{The Ising $Y$-$\nabla$ and $\nabla$-$Y$ transformations
of Ref.~\onlinecite{colbourn1995} each contain 3 square roots, suggesting that there are 8 solutions, but only 2 of these are valid.  In general the couplings may be complex, and choosing principal values for the roots does not guarantee a valid solution: $\delta\beta\gamma=D\sqrt{AC/BD}\sqrt{AB/CD}$ is not necessarily equal to $A$ as required.  The transformations in Ref.~\onlinecite{loh2006} are written such that principal values give a valid solution.}  For the $\nabla$-$Y$ transformation this reflects the $Z_2$ symmetry of the action: for a given $Y$, one can obtain an equivalent $Y$ by flipping the signs of $\sigma_0$, $K_{01}$, $K_{02}$, and $K_{03}$ simultaneously.  There is no clear physical interpretation of the double-valuedness of the $Y$-$\nabla$ transformation.

Setting one or more of the $k_{i}$ to 1 in Eqs.~\ref{e:isingynablaalgebraic} gives the star-mesh formulas for $N=2,1,0$:
	\begin{align}
	N=2 :  \quad&
	k_{12} = \frac{k_1 + k_2}{1 + k_1k_2}, \quad
	z_0 = 1 + k_1k_2
	\\
	N=1 :  \quad&
	z_0 = 1 + k_1
	\\
	N=0 :  \quad&
	z_0 = 2
	\end{align}
The $N=0$ case (integrating out a free spin) simply increases the constant `free energy' term by 
$\ln 2$, 
{\em i.e.}, by one bit.

The special cases below are useful for computer implementation.  They are equivalent to Eqs.~\eqref{e:conductancespecialcases}, with $G$ replaced by $K$:
	\begin{align}
	\mathtt{YDelta} (0, k_1, k_2) &= (1, k_2, k_1) \nonumber\\
	\mathtt{DeltaY} (1, k_1, k_2) &= (0, k_2, k_1) \nonumber\\
	\mathtt{DeltaY} (0, k_1, k_2) &= (k_1 k_2, 0, 0) .
	\label{e:isingspecialcases}
	\end{align}

\myhrulefill\section{The bond propagation procedure}\label{s:bondprop}
The transformations derived in the previous section can be used to \emph{reduce} a network (i.e., eliminate all or nearly all its nodes) and hence to calculate the partition function or effective coupling(s).  
Although an inefficient method, 
gaussian models can be reduced simply by eliminating the nodes one by one, in any order, using the $N$-star-mesh transformation, Eq.~\eqref{e:gaussianstarmesh}.  
In contrast, for Ising models a general star-mesh transformation does not exist.  However, certain networks can be reduced using only $N=3,2,1,0$ star-mesh and mesh-star transformations; these are known as $Y\nabla Y$-reducible networks.  The bond-propagation algorithm is an efficient example of $Y\nabla Y$-reduction.

\begin{figure}
{\centering
\subfigure[A single bond propagation move]
{\resizebox*{0.92\columnwidth}{!}{\includegraphics{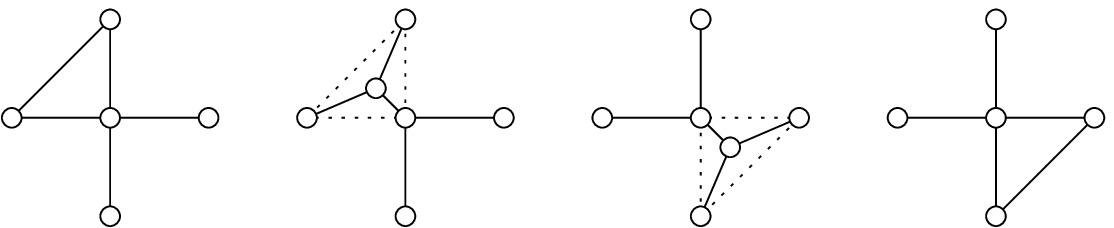}}}
\\
\subfigure[Lattice reduction]
{\resizebox*{0.92\columnwidth}{!}{\includegraphics{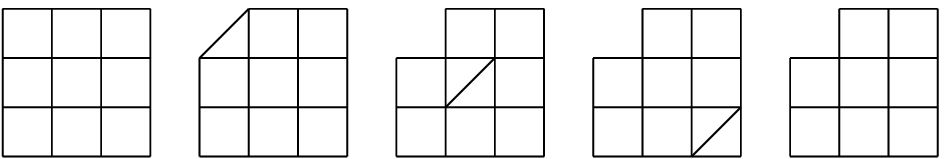}}}
\par}
\caption{
The bond-propagation algorithm in the original form 
invented by Frank and Lobb.  
(a) A single bond propagation move consists of a $Y$-$\nabla$ transformation
followed by a $\nabla$-$Y$ transformation. 
(b) To reduce a finite square lattice, first a corner node is eliminated by series reduction,
producing the first diagonal bond, which is then propagated diagonally down
and to the right until it is ``absorbed'' at the opposite edge.  
Other corner nodes are eliminated in the same way.
}
\label{f:bondproporiginal}
\end{figure}

\subsection{The original bond-propagation algorithm}
Gaussian models and Ising models on square lattices
\footnote{We use `lattice' to mean a `finite lattice'---a periodic structure that has been truncated at a closed bounding surface.}
 can be reduced by the bond propagation procedure \cite{lobb,loh2006}  depicted in Fig.~\ref{f:bondproporiginal}.
 A single bond propagation move is illustrated in Fig.~\ref{f:bondproporiginal}(a).
First, a $\nabla$-$Y$ transformation is performed on the upper left $\nabla$
of Fig.~\ref{f:bondproporiginal}(a).  This introduces one new node.  
The new node is now effectively shifted so as to replace the node to its lower right,
a ``move'' which does not change the topology of the network.  
Then, a $Y$-$\nabla$ transformation is used to convert the 
lower right $Y$ into a $\nabla$, removing a node.   
In this way, any diagonal bond can be
``propagated'' into a diagonally adjacent plaquette.  
The prescription originally implemented in Ref.~\onlinecite{lobb}
is shown in Fig.~\ref{f:bondproporiginal}(b).  
Starting from the upper left corner in Fig.~\ref{f:bondproporiginal}(b),
a series reduction is used to convert the corner into a diagonal bond.  
Then using successive bond propagation moves, 
this diagonal bond can be moved diagonally down and to the right until it annihilates at an edge with open boundary conditions.  

For resistor networks, repeated bond propagation moves reduce the network to a single string of conductors in series, which is easily reducible to one effective conductance.
For an Ising model, the effective coupling is related to the correlation function between corner spins: $\mean{\sigma_{11} \sigma_{NN}} = \tanh K_{11,NN}$.
One may also wish to calculate the partition function in the Ising  model; in order to do this, one must collect the contributions to the partition function during every transformation of the network.  At the end of the calculation, after the last spin or node is eliminated, one has the free energy; the penultimate step gives the effective Ising coupling or effective resistance.

The number of operations necessary to accomplish this is of the order of $O(L^3)$.  Ref.~\onlinecite{lobb} showed empirically that by taking advantage of the early termination of bond propagation on dilute lattices, it is possible to achieve $O(L^2\ln L)$ computational time scaling near the percolation threshold.

\subsection{Useful variants}\label{sec:variants}
\begin{figure}
\centering\includegraphics[width=0.9\columnwidth]{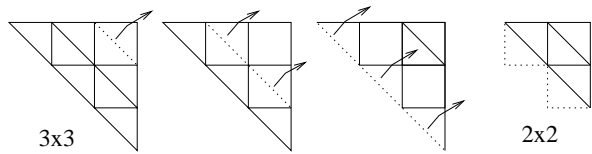}
\caption{Reduction of a triangular lattice of side $L$ to one of side $L-1$ using $\frac{1}{2}L^2$ bond propagations and $L$ series reductions.
\label{f:bondprop4}}
\end{figure}

Figure \ref{f:bondprop4} illustrates a variant of the algorithm for a (finite) triangular lattice.
Here, by embedding the triangular lattice in a square lattice, we see that one third of
the bonds can be considered ``diagonal'' to begin with.  If these  diagonal bonds are only propagated out when it is necessary to make space for other diagonal bonds, then
this procedure reduces the size of the lattice by one unit in every direction. 
Alternatively, the diagonal bonds can all be propagated out, and then
the BPA can proceed as with any square lattice.  

\begin{figure}
\centering
{\resizebox*{0.2\columnwidth}{!}{\includegraphics{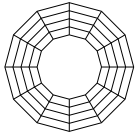}}}
\caption{A cylinder, when viewed in perspective from one end, is equivalent to 
an annulus, which is planar and hence $Y\nabla Y$-reducible.
\label{f:cylinder_annulus}}
\end{figure}

Any planar network can be embedded in a square or triangular lattice (e.g., by inserting `opens' and `shorts' which do not change the partition function), to which the BPA can then be applied.
A network with cylindrical boundary conditions is equivalent to an annular network, which is planar (see Fig.~\ref{f:cylinder_annulus}).  Networks with toroidal boundary conditions, however, are non-planar.

\begin{figure} 
\psfrag{K1}{$K^\text{e}_1$}
\psfrag{K2}{$K^\text{e}_2$}
\psfrag{K3}{$K^\text{e}_3$}
\centering\includegraphics[width=0.43\textwidth]{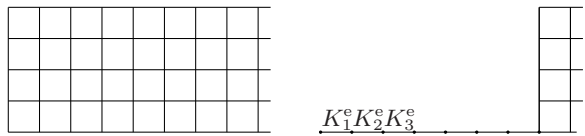}
\caption{Bond propagation for strips.  From the asymptotic distribution of $K^\text{e}_n$ one can infer quantities such as the asymptotic effective resistance per unit length as a function of strip width $M$ (for a random resistor network), or the correlation length $\xi_M$ for a random-bond Ising model of width $M$ such that $\mean{\sigma_{1,1} \sigma_{L,1}} \sim e^{-L/\xi_M}$ for large $L$. \label{f:strip}}
\end{figure}

Self-averaging quantities can be calculated efficiently in a strip geometry.
For the rectangular strip in Fig.~\ref{f:strip}, use a series reduction on  the upper left corner
to produce the first diagonal bond, which is propagated down and to the right
until it is removed from the system.  
This creates two new corners in the upper left, to which the same procedure can be applied.
As corners are eliminated, this leaves a string of bonds with couplings $K_1^e$, $K_2^e$, and so on. 
For sufficiently large $n$, the couplings $K_n^e, K_{n+1}^e$, etc., come from the same distribution.  These couplings may be used to determine the spin correlation length $\xi$.
In this way, the BPA is easily adapted for the calculation of self-averaging quantities in strip geometries (Fig.~\ref{f:strip}).  For a rectangular $L \times M$ strip, its time requirement scales as $O(LM^2)$.

\begin{figure}
\centering\includegraphics[width=1.0\columnwidth]{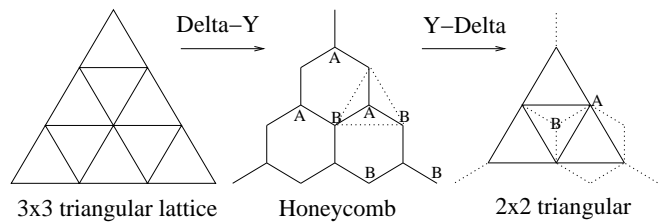}
\caption{Highly parallel algorithm for reduction of a triangular lattice.
Starting with a triangular lattice of $A$ sites, perform $\nabla$-$Y$ transformations to insert $B$ sites.  On the resulting honeycomb lattice, perform $Y$-$\nabla$ transformations to eliminate all the $A$ sites.  This gives a triangular lattice that is one unit smaller in every direction.  
\label{f:bondprop5}}
\end{figure}

Instead of using the BPA $Y\nabla Y$ reduction procedure,
efficient {\em parallel algorithms} may be developed by direct 
$Y\nabla Y$ reduction on every plaquette simultaneously.
The algorithm shown in Fig.~\ref{f:bondprop5} represents the same arithmetic operations performed in a different order.  
Starting with a triangular lattice, one performs a $\nabla$-$Y$ transformation on every upward-pointing $\nabla$, thus producing a honeycomb lattice.  One then performs a $Y$-$\nabla$ transformation on every downward-pointing $Y$, producing a triangular lattice one unit smaller than the original.

\begin{figure}
\centering\includegraphics[width=0.4\textwidth]{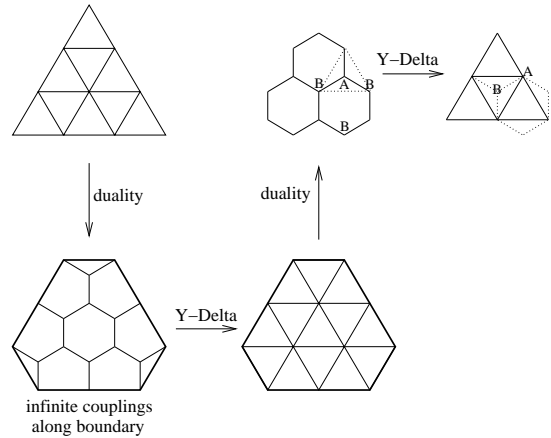}
\caption{A variant of Fig.~\ref{f:bondprop5} using duality and $Y$-$\nabla$ transformations only.  Thick lines represent infinite-strength bonds. 
\label{f:bondprop7}}
\end{figure}

Using the `cyclic change of lattices' introduced by Syozi\cite{syozi}, it is even possible to devise an algorithm involving only $Y$-$\nabla$ and duality transformations, as in Fig.~\ref{f:bondprop7}.  The dual of a planar network is derived by drawing a dual bond cutting across each bond of the original network, with a bond strength given by the duality relation $\mathtt{dual}(G)=\frac{1}{G}$ (for resistors) or $\mathtt{dual}(k)=\frac{1-k}{1+k}$ (for Ising bonds).

The latter two algorithms for $Y\nabla Y$ reduction are very suitable for parallelization.  In Figs.~\ref{f:bondprop5} and \ref{f:bondprop7}, the $O(L^3)$ operations can be distributed among $L^2$ processors so that the entire network can be reduced in only $O(L)$ time.  

\subsection{Reduction of networks with external terminals}
\begin{figure}
\subfigure[]{\includegraphics{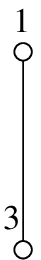}\label{f:irred2}}  
\hspace{10mm}
\subfigure[]{\includegraphics{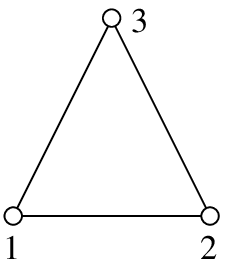}\label{f:irred3}}  
\hspace{10mm}
\subfigure[]{\includegraphics{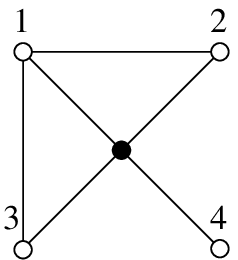}\label{f:irred4}}
\caption{Irreducible $2$-, $3$-, and $4$-terminal networks.\label{f:irrednetworks}}
\end{figure}

\begin{figure}
\centering\includegraphics[width=0.25\textwidth]{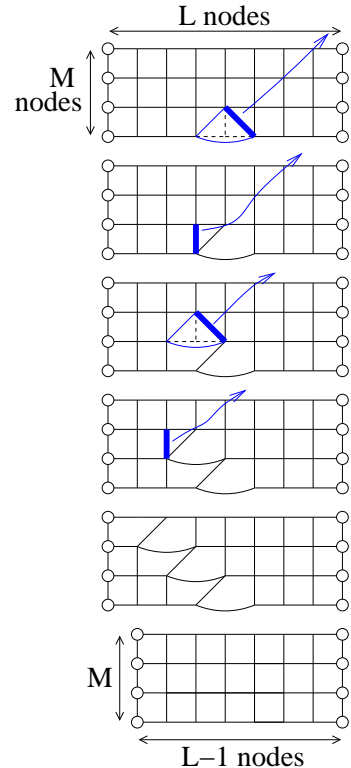}
\caption{`Contraction' of an $L\times M$ rectangular strip preserving nodes along short edges, where $L>2M$.  Dotted lines denote bonds that are removed by $Y$-$\nabla$ transformations; bold lines indicated bonds that are propagated out.  Each curved arrow indicates a sequence of bond propagations. 
\label{f:stripreduction}}
\end{figure}

When the BPA is used to compute the effective couplings between an even number of `external' or `terminal' spins ($2, 4, 6, \dotsc$), one must take care \emph{not} to eliminate these terminal spins.
Given any two-terminal planar network, it is always possible to integrate out all the `internal' spins, reducing the network to a single bond between the two terminals (Fig.~\ref{f:irred2}).
However, for a general 4-terminal planar network, it is not possible to reduce the network beyond the stage illustrated in Fig.~\ref{f:irred4}, which has one internal spin.  It may be interesting to find irreducible networks with larger numbers of terminals (see, e.g., Ref.~\onlinecite{archdeacon2000}).
A rectangular $L\times M$ strip ($L>2M$) with terminals all along its short edges can be reduced to a $2M\times M$ strip by successive applications of the procedure in Fig.~\ref{f:stripreduction}; this may be thought of as a decimation procedure which preserves the form of the Ising Hamiltonian.

For a triangular network with one terminal at each corner, the algorithm in Fig.~\ref{f:bondprop5} leads to an irreducible three-terminal network (Fig.~\ref{f:irred3}), from which the effective two-terminal couplings $K_{23}$, $K_{31}$, $K_{12}$ may easily be found.  For uncorrelated random-bond models, this approach produces effective couplings for three disorder realizations for the price of one.


\myhrulefill\section{Bond propagation for calculating derivatives of the partition function}\label{s:jeremy}
The algorithm described above (and its variants) allows us to calculate the numerical value of the partition function $Z(\beta)$ at one value of $\beta$.  To compute other thermodynamic quantities such as the mean energy $U$, one approach is to calculate $Z$ at closely spaced 
values of $\beta$ and perform numerical differentiation.  This is not appealing because 
while the algorithm produces near-machine-precision values for $Z$,
this would yield only approximate values for $U$.
We now show that it is in fact possible to extend the bond-propagation algorithm to compute derivatives of $Z$ \emph{directly}, i.e., without numerical differentiation.  

The quantity 
$U = -\frac{d}{d\beta}{\rm ln}Z(\beta)$
can be computed from
	\begin{align}
	-\beta ZU 
	&= \sum_{\{\sigma\}}
		S(\{\sigma\})
		e^{ S(\{\sigma\}) }
		\nonumber\\
	&= \sum_{\{\sigma\}}
		\Big(  \sum_{\mean{ij}} K_{ij} \sigma_i \sigma_j  \Big)
		\exp
		\sum_{\mean{ij}} K_{ij} \sigma_i \sigma_j
	\label{e:betazu}
	\end{align}
In order to use the BPA to calculate this quantity directly, we develop
$Y$-$\nabla$ and $\nabla$-$Y$ transformations which preserve this quantity.
First, consider a `$Y$' with terminal spins $\sigma_1$, $\sigma_2$, $\sigma_3$ and central spin $\sigma_0$.  
In fact, we can solve a more general expression, in which the bond strengths in the prefactor
can differ from those in the exponent, such that 
the `$Y$' is described by 8 parameters $A$, $L_{01}$, $L_{02}$, $L_{03}$ and  $F$, $K_{01}$, $K_{02}$, $K_{03}$:
\begin{widetext}
	\begin{align}
	C 
	&=\sum_{[\sigma_{1,2,3,\dotsc}]} \sum_{\sigma_0}
		\Big(
				c_\text{ext}
			+	A
			+ L_{01} \sigma_1 \sigma_0
			+ L_{02} \sigma_2 \sigma_0 
			+ L_{03} \sigma_3 \sigma_0 		
		\Big)
		e^{ 
				S_\text{ext}
			+ F
			+ K_{01} \sigma_1 \sigma_0
			+ K_{02} \sigma_2 \sigma_0 
			+ K_{03} \sigma_3 \sigma_0 
		}.
	\end{align}
In the above expression, the quadratic terms in the prefactor and exponent not involving $\sigma_0$ have been collected in the functions $c_\text{ext}(\sigma_1,\sigma_2,\dotsc)$ and $S_\text{ext}(\sigma_1,\sigma_2,\dotsc)$.  Factorizing and evaluating the sum over $\sigma_0$,
	\begin{align}
	C &=\sum_{[\sigma_{1,2,3,\dotsc}]} 
		e^{S_\text{ext} + F}
	\bigg\{
		\sum_{\sigma_0}
		\big(
				c_\text{ext}
			+	A
		\big)
		e^{
			  K_{01} \sigma_1 \sigma_0
			+ K_{02} \sigma_2 \sigma_0 
			+ K_{03} \sigma_3 \sigma_0 
		}
		\nonumber\\&~~~~~~~~~~~~~~~~{}
		+
		\sum_{\sigma_0}
		\big(
			  L_{01} \sigma_1 \sigma_0
			+ L_{02} \sigma_2 \sigma_0 
			+ L_{03} \sigma_3 \sigma_0 		
		\big)
		e^{
			  K_{01} \sigma_1 \sigma_0
			+ K_{02} \sigma_2 \sigma_0 
			+ K_{03} \sigma_3 \sigma_0 
		}
	\bigg\}
		\nonumber\\	
	&=\sum_{[\sigma_{1,2,3,\dotsc}]} 
		e^{S_\text{ext} + F}
	\bigg\{
									\underbrace{
		2\big(
				c_\text{ext}
			+ A
		\big) 
		\cosh \big(
			  K_{01} \sigma_1
			+ K_{02} \sigma_2
			+ K_{03} \sigma_3
		\big)
														}_\text{1st term}
		\nonumber\\&~~~~~~~~~~~~~~~~{}
		+
									\underbrace{
		2\big(
			  L_{01} \sigma_1
			+ L_{02} \sigma_2
			+ L_{03} \sigma_3
		\big)
		\sinh \big(
			  K_{01} \sigma_1
			+ K_{02} \sigma_2
			+ K_{03} \sigma_3
		\big)
														}_\text{2nd term}
	\bigg\}
	.
	\label{e:zu_y}
	\end{align}
We wish to find the new parameters $A'$, $L'_{23}$, $L'_{31}$, $L'_{12}$ and  $F'$, $K'_{23}$, $K'_{31}$, $K'_{12}$ of the equivalent $\nabla$ such that
	\begin{align}
	C
	&=\sum_{[\sigma_{1,2,3,\dotsc}]}
		\big(
				c_\text{ext}
			+	A'
			+	L'_{23} \sigma_2 \sigma_3
			+	L'_{31} \sigma_3 \sigma_1
			+	L'_{12} \sigma_1 \sigma_2
		\big)
		e^{
				S_\text{ext}
			+ F'
			+	K'_{23} \sigma_2 \sigma_3
			+	K'_{31} \sigma_3 \sigma_1
			+	K'_{12} \sigma_1 \sigma_2
		}
		\nonumber\\
	&=\sum_{[\sigma_{1,2,3,\dotsc}]}
		e^{S_\text{ext} + F}
	\bigg\{
									\underbrace{
		\big(
				c_\text{ext}
			+	A
		\big)
		e^{
				\delta F
			+	K'_{23} \sigma_2 \sigma_3
			+	K'_{31} \sigma_3 \sigma_1
			+	K'_{12} \sigma_1 \sigma_2
		}
														}_\text{1st term}
		\nonumber\\&~~~~~~~~~~~~~~~~~~~~~~~~~~~~~~~~~~~~~~~~~~~~{}
		+
									\underbrace{
		\big(
				\delta A
			+ L'_{23} \sigma_2 \sigma_3
			+	L'_{31} \sigma_3 \sigma_1
			+	L'_{12} \sigma_1 \sigma_2
		\big)
		e^{
				\delta F
			+	K'_{23} \sigma_2 \sigma_3
			+	K'_{31} \sigma_3 \sigma_1
			+	K'_{12} \sigma_1 \sigma_2
		}
														}_\text{2nd term}
	\bigg\}
	\label{e:zu_nabla}
	\end{align}
where $\delta A=A'-A$ and $\delta F=F'-F$.
Comparing Eqs.~\eqref{e:zu_y} and \eqref{e:zu_nabla} suggests equating the two terms inside the braces separately.  Equating the first term leads to
	\begin{align}
	&
		2\cosh \big(
			  K_{01} \sigma_1
			+ K_{02} \sigma_2
			+ K_{03} \sigma_3
		\big)
	=
		e^{
				\delta F
			+	K'_{23} \sigma_2 \sigma_3
			+	K'_{31} \sigma_3 \sigma_1
			+	K'_{12} \sigma_1 \sigma_2
		}
	\qquad\forall \sigma_1,\sigma_2,\sigma_3
	,
	\end{align}
which is identical to Eq.~\eqref{e:equatingfirstterms}.  Equating the second terms in the braces in Eqs.~\eqref{e:zu_y} and \eqref{e:zu_nabla} gives
	\begin{align}
	&
		2\big(
			  L_{01} \sigma_1
			+ L_{02} \sigma_2
			+ L_{03} \sigma_3
		\big)
		\sinh \big(
			  K_{01} \sigma_1
			+ K_{02} \sigma_2
			+ K_{03} \sigma_3
		\big)
	\nonumber\\&=
		\big(
				\delta A
			+ L'_{23} \sigma_2 \sigma_3
			+	L'_{31} \sigma_3 \sigma_1
			+	L'_{12} \sigma_1 \sigma_2
		\big)
		e^{
				\delta F
			+	K'_{23} \sigma_2 \sigma_3
			+	K'_{31} \sigma_3 \sigma_1
			+	K'_{12} \sigma_1 \sigma_2
		}
	\qquad\forall \sigma_1,\sigma_2,\sigma_3.
	\label{e:equatingsecondterms}
	\end{align}
We can now use  Eq.~\eqref{e:equatingfirstterms} to rewrite  Eq.~\eqref{e:equatingsecondterms} as
	\begin{align}
	&
		2\big(
			  L_{01} \sigma_1
			+ L_{02} \sigma_2
			+ L_{03} \sigma_3
		\big)
		\sinh \big(
			  K_{01} \sigma_1
			+ K_{02} \sigma_2
			+ K_{03} \sigma_3
		\big)
	\nonumber\\&=
		\big(
				\delta A
			+ L'_{23} \sigma_2 \sigma_3
			+	L'_{31} \sigma_3 \sigma_1
			+	L'_{12} \sigma_1 \sigma_2
		\big)
		2\cosh \big(
			  K_{01} \sigma_1
			+ K_{02} \sigma_2
			+ K_{03} \sigma_3
		\big)
	\qquad\forall \sigma_1,\sigma_2,\sigma_3
	\\
	\therefore\qquad&
		\big(
				\delta A
			+ L'_{23} \sigma_2 \sigma_3
			+	L'_{31} \sigma_3 \sigma_1
			+	L'_{12} \sigma_1 \sigma_2
		\big)
	=
		\big(
			  L_{01} \sigma_1
			+ L_{02} \sigma_2
			+ L_{03} \sigma_3
		\big)
		\tanh \big(
			  K_{01} \sigma_1
			+ K_{02} \sigma_2
			+ K_{03} \sigma_3
		\big)	
	\\
	\therefore\qquad&
	\begin{cases}
	\delta A + L'_{23} + L'_{31} + L'_{12}	= u_0=(+L_{01}+L_{02}+L_{03} ) 
	\tanh (+K_{01}+K_{02}+K_{03}) 	\\
	\delta A + L'_{23} - L'_{31} - L'_{12}	= u_1=(-L_{01}+L_{02}+L_{03} ) 
	 \tanh (-K_{01}+K_{02}+K_{03}) 	\quad\text{(cycl.)}\\
	\end{cases}
	\label{e:blahblah}
	\\
	\therefore\qquad&
	\begin{cases}
	\delta A = (u_0 + u_1 + u_2 + u_3)/4 \\
	L'_{23} = (u_0 + u_1 - u_2 - u_3)/4 \quad\text{(cycl.)}\\
	\end{cases}
	\label{e:jeremyynabla}
	\end{align}
where $u_i$ are functions of $L_{0i}$ and $K_{0i}$ as defined above.
	\end{widetext}

This $Y$-$\nabla$ transformation can be inverted to give a $\nabla$-$Y$ transformation.  First, compute $K_{01}$, $K_{02}$, $K_{03}$ using	Eq.\eqref{e:isingnablayalgebraic}.  Then let 
	\begin{align}
	\begin{cases}
	\delta A = \frac{ (c_0-c_1+c_2+c_3) L'_{23} + cycl. }{ -c_0+c_1+c_2+c_3 } ,  \\
	L_{01} = \frac{ (c_0-c_1)(c_2+c_3) L'_{23} + (c_1c_2+c_0c_3) L'_{31} + (c_3c_1+c_0c_2) L'_{12} }{ -c_0+c_1+c_2+c_3 } 
	\quad\text{(cycl.)}\\
	\end{cases}
	\label{e:jeremynablay}
	\end{align}
where
	\begin{align}
	\begin{cases}
	c_0 = \coth (K_{01} + K_{02} + K_{03}) \\
	c_1 = \coth (-K_{01} + K_{02} + K_{03}) 
	\quad\text{(cycl.)} . \\
	\end{cases}
	\end{align}
Eqs.~\eqref{e:isingynablaalgebraic},\eqref{e:isingnablayalgebraic},\eqref{e:jeremyynabla},\eqref{e:jeremynablay} can be used in the bond-propagation algorithm (or any other $Y\nabla Y$ reduction scheme) to compute the mean energy $U$ of an Ising network directly, without numerical differentiation.  
Note that since the transformations which preserve $Z$ are not the same
as those which preserve its derivative $Z^{\prime}$, 
obtaining  $U = -Z^{\prime}/Z$ requires two separate bond propagation calculations:
one to calculate $Z$; the other to calculate $Z^{\prime}$.
It may be possible to calculate the heat capacity and other higher derivatives of $Z$ in a similar manner.

\myhrulefill\section{Discussion}\label{s:discussion}

We have found that bond propagation generally gives answers to high accuracy (close to machine precision), provided that one uses a careful
implementation that avoids overflow and underflow and minimizes roundoff error, especially in the case of large systems, low temperatures, or frustration.
We discuss specific cases below. 

\subsection{Ferromagnetic models}
The algorithm works rather straightforwardly for ferromagnetic Ising models where all the $K_{ij}$ are positive.  The $Y$-$\nabla$ and $\nabla$-$Y$ transformations should be implemented in a way that minimizes roundoff error, as in Eqs.~\eqref{e:isingynablaalgebraic} and \eqref{e:isingnablayalgebraic}.
For large systems in the ferromagnetic phase $(L \gtrsim 10^3)$, the partition function $Z$ may cause floating-point overflow, which can easily be dealt with by storing $F=\ln Z$ instead.  The couplings themselves $k_{ij}$ may also underflow, in which case it may be necessary to use an alternative representation to store the couplings and perform the transformations, or to change the order of the bond propagations.

\subsection{Dilute models}
Applying bond propagation to dilute models containing zero couplings may generate infinite couplings, as pointed out by Frank-Lobb.  We have confirmed that this can be successfully dealt with using Eqs.~\eqref{e:conductancespecialcases} or \eqref{e:isingspecialcases}.

\subsection{Frustration}
All couplings $K_{ij}$ in a physical Ising model must be real-valued.
The $Y$-$\nabla$ transformation, Eq.~\eqref{e:isingynablaalgebraic}, preserves this property, but the $\nabla$-$Y$ transformation, Eq.~\eqref{e:isingnablayalgebraic}, can generate complex-valued couplings.  This happens if the argument of the square root is negative ($v^2=y_1{}^2 - x_1{}^2<0$), i.e.,
	\begin{align}
&
    (1+k_{23}k_{31}+k_{31}k_{12}+k_{12}k_{23}) \nonumber\\&{}\times
    (1+k_{23}k_{31}-k_{31}k_{12}-k_{12}k_{23}) \nonumber\\&{}\times
    (1-k_{23}k_{31}+k_{31}k_{12}-k_{12}k_{23}) \nonumber\\&{}\times
    (1-k_{23}k_{31}-k_{31}k_{12}+k_{12}k_{23})
    < 0 .
	\end{align}
This ``frustration inequality'' can be taken as the definition of a ``frustrated $\nabla$''.\footnote{There is an interesting analogy in geometry: the area of a triangle, $A=\frac{1}{4}\sqrt{(a+b+c)(b+c-a)(c+a-b)(a+b-c)}$, must satisfy $A^2\geq 0$, which leads to ``triangle inequalities''---geometrical constraints on the lengths of the sides of a triangle.}  Frustration occurs in $\nabla$'s where all 3 couplings have similar magnitudes and 1 or 3 or them are antiferromagnetic ($K_{ij}<0$, $k_{ij}>1$).  The BPA still works for frustrated systems provided that complex arithmetic is used.  Despite the occurrence of complex intermediate couplings, we have found that the value of the partition function emerging from the entire calculation is real, as expected, with a small imaginary part arising from roundoff error.

The models typically studied in the literature are random-bond Ising models with bond strengths picked independently from the same distribution, $P(K)$.  The two most common models are the `binary' ($\pm K_0$) RBIM, with $P(K_{ij})=(1-p)\delta(K_{ij}-K_0) + p\delta(K_{ij}+K_0)$ where $p$ is the concentration of antiferromagnetic couplings, and the Gaussian RBIM, with $P(K_{ij}) \propto \exp -\frac{1}{2} \left(\frac{K_{ij}-K_0}{\sigma_K}\right)^2$.  

The BPA works well for the Gaussian RBIM.
For the binary RBIM, certain disorder realizations contain perfectly frustrated plaquettes; these cause the BPA to generate infinite and zero couplings, which then lead to indeterminate results.  
This is due to singularities in the function represented by the bond propagation procedure (the composition of all the $Y$-$\nabla$ and $\nabla$-$Y$ transformations).  These can be avoided using perturbations away from perfect frustration, including roundoff error itself.   
Such perturbations may be relevant in a renormalization-group sense, changing the universality class; if the scaling flow is weak, it may still be possible to obtain accurate results this way.  Introducing perturbations this way is acceptable as long as the systematic error it produces does not exceed the random statistical error from disorder averaging, which is an inevitable part of numerical calculations on RBIMs.	
At very low temperatures roundoff error may become an issue for both types of RBIM.

\subsection{Comparisons with other algorithms}
Gaussian models (including resistor networks) possess $N$-star-mesh transformations for all $N\geq 1$, so there are a large number of algorithms for reducing them.  Below is a comparison of the number of multiplications ($\times$) and divisions ($\div$) required to perform various algorithms on Gaussian models ({\em e.g.}, resistor networks) on an $L\times L$ lattice:
\begin{itemize}
\item Cramer's rule: $O(L^2!)~(\times)$, $O(1)~(\div)$
\item Full Gauss elimination: $O(L^6)~(\times)$ , $O(L^2)~(\div)$
\item Transfer matrices\cite{derrida1982,derrida1982b}: $O(L^4)~(\times)$ , $O(L^3)~(\div)$ 
\item Banded Gauss elim.: $O(L^4)~(\times)$ , $O(L^2)~(\div)$
\item Nested dissection\cite{george1973,lipton1979}: $O(L^3)~(\times)$ , $O(L^2)~(\div)$ 
\item Bond propagation\cite{lobb}: $O(L^3)~(\times)$ , $O(L^3)~(\div)$ 
\end{itemize}
The last two methods are the fastest, both taking $O(L^3)$ time overall.
The nested dissection algorithm developed by George\cite{george1973} and 
by Lipton {\em et al.}\cite{lipton1979} is a hierarchical ``divide-and-conquer'' method that exploits prior knowledge about the connectivity of the network.  It is quite a general idea and it can be used with doubly periodic boundary conditions, as well as for lattices in dimensions $d>2$, for which the time requirement is $O(L^{3d-3})$.
Its drawback is the amount of bookkeeping required.  
On the other hand, bond propagation, while slightly more costly in terms of divisions,
is simpler to implement and especially to parallelize. 
(See Sec.~\ref{sec:variants}.)
For dilute networks near the percolation threshold, bond propagation 
is even faster, 
taking $O(L^2 {\rm ln}L)$ time, because many propagating bonds terminate early. 
In the context of Gaussian models, bond propagation can be viewed as an efficient linear algebra method for reducing certain pentadiagonal matrices while preserving their sparsity. 

For Ising models, polynomial-time methods are only known to exist in two dimensions.  Below are operation counts including addition ($+$), subtraction ($-$), multiplication ($\times$), division ($\div$), and square roots ($\sqrt{}$) for selected methods:
\begin{itemize}
\item Transfer matrices in the spin basis\cite{morgenstern1979}: $O(2^L)$ operations $(+\times)$ and memory; can be generalized to higher-dimensional lattices
\item Fermion network method\cite{merz}: $O(L^4)$ operations $(+\times)$
\item Pfaffian elimination with nested dissection\cite{chineseremainder}: $O(L^3)$ operations $(+\times\div)$
\item $Y\nabla Y$ reduction by Feo-Provan algorithm\cite{colbourn1995}: $O(L^4)$ operations $(+-\times\div\sqrt{})$ for general graphs
\item $Y\nabla Y$ reduction by bond propagation\cite{loh2006}: $O(L^3)$ operations $(+-\times\div\sqrt{})$
\end{itemize}
Due to the lack of a general $N$-star-mesh transformation for Ising models, nested dissection cannot be applied directly to the Ising network.  Rather, one must first map the Ising problem to a dimer problem on the Kasteleyn `terminal lattice' which has 4 nodes for each Ising spin.  In two dimensions, the dimer problem maps to a Pfaffian elimination problem, to which nested dissection can then be applied.
In contrast, the bond propagation algorithm can be performed directly in the Ising network representation,
with no need to map to dimers or fermion models. 

\subsection{Limitations}  \label{s:othermodels}
$Y\nabla Y$ reduction techniques (of which the bond-propagation algorithm is one) for statistical mechanics models rely on two criteria: (a) $Y\nabla Y$-reducibility of the graph on which the model is defined and (b) the existence of $Y$-$\nabla$ and $\nabla$-$Y$ transformations that preserve 
some aspect of the model at hand. 

We first discuss criterion (a).  
The Robertson-Seymour theory of forbidden minors and obstruction sets (see Ref.~\onlinecite{robertson2004} and references therein)
implies the following relationships among various sets of graphs: 
	\begin{align}
	&\{\text{Forests} \}
	\nonumber\\& \subset \{ \text{Outerplanar graphs} \}
	\nonumber\\& \subset \{ \text{Series-parallel-reducible graphs} \}
	\nonumber\\&	\subset \{ \text{Planar graphs} \}
	\nonumber\\&	\subset \{ \text{$Y\nabla Y$-reducible graphs} \}
	\nonumber\\&	\subset \{ \text{Linklessly embeddable graphs} \} 
	.
	\end{align}
In particular, this shows that all planar graphs are $Y\nabla Y$-reducible.  
However, it is also true that nearly all `sufficiently large' non-planar graphs with crossing number 2 or greater are not $Y\nabla Y$-reducible.  This means that $Y\nabla Y$ approaches such as bond propagation are unable to completely reduce 3D lattices or even 2D `pyrochlore' lattices (lattices of edge- or corner-sharing tetrahedra).
For example, a large enough 3D network with simple cubic symmetry
has as a  minor the Petersen graph 
$KG_{5,2}$.  Since $KG_{5,2}$ 
is a forbidden minor for $Y \nabla Y$ reduction,
the simple cubic lattice is not amenable to the BPA, nor to any
algorithm based purely on $Y \nabla Y$ reduction.
Since general 3D lattices contain the simple cubic lattice as a minor,
they are not $Y \nabla Y$-reducible either.  
As a further example, although diagonal bonds can be made to propagate on, {\em e.g.,}
 a tiling of 3-space by truncated octahedra, this 3D lattice is not {\em reducible}
 by bond propagation.
Of course, even for networks that are not \emph{completely} $Y\nabla Y$-reducible, it may be useful to perform \emph{partial} $Y\nabla Y$-reduction before resorting to more expensive computational techniques.

We now turn to criterion (b). 
In Secs.~\ref{s:transformations} and \ref{s:jeremy} we derived $Y$-$\nabla$ and $\nabla$-$Y$ transformations for  Gaussian models and Ising models.  These models have the special property that 3-spin interactions are forbidden by symmetry.  For most other models in statistical mechanics, eliminating a 3-coordinated spin generates 3-spin interactions which cannot be described  within the network formalism.
For example, consider the Ising model in the presence of magnetic fields $h_i$ (which may be uniform or random), described by the action
	\begin{align}
	e^{ S(\{\sigma\})  }
	&= \exp \Big( \sum_{\mean{ij}} K_{ij} \sigma_i \sigma_j + \sum_i h_i \sigma_i  \Big) .
	\label{e:rfimaction}
	\end{align}
Let us integrate out the center spin of a $Y$ and find the effective action for the resultant $\nabla$
by equating Boltzmann weights for each of the $8$ configurations of the outer spins ($\sigma_1,\sigma_2,\sigma_3=\pm 1$):
	\begin{align}
	\sum_{\sigma_0} e^{ S(\sigma_0,\sigma_1,\sigma_2,\sigma_3)  }
	&=  e^{ S'(\sigma_1,\sigma_2,\sigma_3)  } .
	\end{align}
If we restrict ourselves to the form of 	
Eq.~\eqref{e:rfimaction}, there are only 7 possible terms:
$S' = F + h_1 \sigma_1 + h_2 \sigma_2 + h_3 \sigma_3 + K_{23} \sigma_2 \sigma_3 + K_{31} \sigma_3 \sigma_1 + K_{12} \sigma_1 \sigma_2$.
Therefore $S'$ must contain a 3-spin interaction $K_{123} \sigma_1 \sigma_2 \sigma_3$, and it cannot be written in the same analytic form as the original action:
 a $Y$-$\nabla$ transformation does not exist.

Table \ref{t:table1} shows a `feasibility study' for $Y-\nabla$ transformations for the $q$-state Potts model, with couplings $K_{ij}$ that tend to align adjacent spins in the same `direction' in spin space,
	\begin{align}
	e^{ S(\{\sigma\})  }
	&= \exp \sum_{\mean{ij}} K_{ij} \delta_{\sigma_i \sigma_j}, 
	~ \sigma_i \in \{1,2,\dotsc,q\} \forall i .
	\label{e:pottsaction}
	\end{align}
a generalized Potts model (where every bond is described by a matrix $\mathbf{P}_{ij}$ representing the weight factors for every combination of directions of two adjacent spins),
	\begin{align}
	e^{ S(\{\sigma\})  }
	&= \exp \sum_{\mean{ij}} P_{ij\sigma_i\sigma_j},
	~ \sigma_i \in \{1,2,\dotsc,q\} \forall i .
	\end{align}
and the $q$-state clock model (where $q\geq 3$),
	\begin{align}
	e^{ S(\{\sigma\})  }
	&= \exp \sum_{\mean{ij}} K_{ij} \cos \tfrac{2\pi}{q} (\sigma_i - \sigma_j).
	\label{e:clockaction}
	\end{align}
For these models there are, again, more equations than parameters, so exact $Y$-$\nabla$  transformations do not exist.

In the limit $q\rightarrow \infty$ the clock model becomes the XY model.  This suggests that $Y$-$\nabla$ transformations do not exist for XY models, nor for other models with continuous spins (\emph{e.g.}, classical Heisenberg models).

\begin{table*} 
\begin{tabular}{c|c|l|l} \hline
Type of model & Distinct equalities & $Y$ parameters & $\nabla$ parameters \\ \hline \hline
Ising  & 4 & 4 $(F,K_{01},K_{02},K_{03})$ & 4 $(F,K_{23},K_{31},K_{12})$ \\
Ising with fields & 8 & 8 $(F,K_{01},K_{02},K_{03},h_1,h_2,h_3,h_0)$ & 7 $(F,K_{23},K_{31},K_{12},h_1,h_2,h_3)$ \\
Potts ($q \geq 3$) & 5 & 4  & 4   \\ 
Potts (generalized, $q\geq 3$) & $q^3$ & $3q^2-2$ & $3q^2-2$ 
\\
 \hline
\end{tabular}
\caption{\label{t:table1}Degrees of freedom of $Y$ and $\nabla$ sub-networks.  
For all cases except the standard Ising model, there are more equations than $\nabla$ parameters, indicating that the effective action contains 3-spin interactions and that exact $Y$-$\nabla$ transformations do not exist.
}
\end{table*}

This paper has focused on statistical mechanics, but $Y\nabla Y$-reduction and the BPA also have important applications in combinatorics and operations research.
Exact $Y$-$\nabla$ and $\nabla$-$Y$ transformations exist 
for the shortest path problem, for the max flow (or min cut) problems,\cite{feo1993} for counting spanning trees and perfect matchings,\cite{colbourn1995}, and even for 2D foams in mechanical equilibrium\cite{mancini2007}.
Such transformations do \emph{not} exist for 
network reliability problems\cite{lehman1963} 
\footnote{The network reliability problem is more commonly known in physics as `bond percolation'.  The BPA can be used to reduce a planar network with a particular configuration of `active' bonds, but it is not possible to automatically average over all bond configurations 
based purely on the bond probabilities.},
nor for equilibrium flow problems with arbitrary potential-flow relations \cite{feo1993} (\emph{i.e.}, electrical circuits composed of elements with arbitrary non-linear current-voltage characteristics).
Thus, the former group of problems can be solved on planar networks in polynomial time, whereas the latter group cannot.
In Table \ref{t:table2}, all of the above problems are classified according to the maximum degree of star-mesh transformations that they support. 

The above examples have been for systems defined on undirected graphs, i.e., `symmetric' bonds.
$Y\nabla Y$-reduction (and the BPA) may also be applicable to \emph{directed} graph problems such as dynamical Ising cellular automata.\cite{odor2004}  
Note that in systems where \emph{exact} $Y$-$\nabla$ transformations do not exist,  \emph{approximate} $Y$-$\nabla$ transformations may still have some use.\cite{lehman1963}

\begin{table}
\begin{tabular}{c|c|c} \hline
Type of model & $N^\text{star-mesh}_\text{max}$ &  $N^\text{mesh-star}_\text{max}$ \\ \hline \hline
Gaussian models,      & &\\
resistor networks,      & &\\
shortest path, &$\infty$ & 3 \\
max flow & & \\
\hline
Ising model        & 3 & 3 \\ 
\hline
Ising with field  & 2 & 3 \\ 
\hline
Potts, clock, &  &  \\
XY, Heisenberg, &  2 &2  \\
network reliability   & & \\ \hline
\end{tabular}
\caption{\label{t:table2} Maximum degree of star-mesh and mesh-star transformations for various models.
}
\end{table}

\myhrulefill\section{Conclusion}
In conclusion, we have shown that the bond propagation algorithm (BPA) may be applied not only
to the partition function and correlation functions in the Ising model, but that it can also be applied
to thermodynamic quantities via transformations which preserve the first derivative of the partition function.
Similar transformations may exist for higher derivatives as well.  
Since the BPA is a form of $Y \nabla Y$ reduction, it is limited to networks which are
$Y \nabla Y$-reducible.  This includes planar graphs, and the BPA may be applied to 
Ising models on planar networks as well as planar resistor networks.  In general, 3D 
lattices are not $Y \nabla Y$-reducible, and so the BPA is not amenable to 
3D resistor networks or Ising models in 3D.  

\myhrulefill\section{Acknowledgments}
It is a pleasure to thank R.~Fisch,  E.~H.~Goins, C.~J.~Lobb,  and F.~Merz for helpful discussions.
This work was supported by Purdue University and Research Corporation (YLL) and by the Purdue Research Foundation (EWC).   EWC is a Cottrell Scholar of Research Corporation.

\bibliographystyle{forprl}
\bibliography{rbim.prb}

\end{document}